\documentstyle[prb,aps,preprint]{revtex}
\begin{document}
\draft
\title{Chiral Symmetry and Collective Excitations in p-wave, d-wave and f-wave Superconductors}
\author{Tadafumi Ohsaku}
\address{Research Center for Nuclear Physics, Osaka University, Ibaraki, Osaka, Japan.}

\date{\today}
\maketitle

\newcommand{\bmx}{\mbox{\boldmath $x$}}
\newcommand{\bmy}{\mbox{\boldmath $y$}}
\newcommand{\bmk}{\mbox{\boldmath $k$}}
\newcommand{\bmp}{\mbox{\boldmath $p$}}
\newcommand{\bmq}{\mbox{\boldmath $q$}}
\newcommand{\bmP}{\mbox{\boldmath $P$}}  
\newcommand{\kfey}{\ooalign{\hfil/\hfil\crcr$k$}}
\newcommand{\pfey}{\ooalign{\hfil/\hfil\crcr$p$}}
\newcommand{\qfey}{\ooalign{\hfil/\hfil\crcr$q$}}
\newcommand{\Deltafey}{\ooalign{\hfil/\hfil\crcr$\Delta$}}  
\def\sech{\mathop{\rm sech}\nolimits}

\begin{abstract}

We discuss the origin of charge density wave ( CDW ) and spin density wave ( SDW ) 
in p-wave, d-wave and f-wave superconductors. 
To describe the low-energy quasiparticle excitation of p-wave case, 
we introduce a two- ( one for time and one for space ) dimensional massless Dirac model. 
After the non-Abelian bosonization is performed, 
the charge and spin density waves emerge from the model. 
By using this scheme, we try to explain the characteristic aspect of phase diagrams of various compounds, 
oxides and organic superconductors.
The purpose of this paper is to make an argument that the dimensionality of the nodal excitation
in superconductors plays an important role in the determination of the structure
of the phase diagram.

\end{abstract}

\vspace{20mm}

Motivated by recent experimental discoveries of the coexistence of 
antiferromagnetic and superconducting orders~[1], 
Franz and Tesanovic, and independently Herbut found that the chiral symmetry and its dynamical breaking 
( dynamical chiral symmetry breaking, ${\rm D}\chi{\rm SB}$ ) 
is realized in d-wave copper oxide superconductors~[2]. 
They introduced a four-component Dirac field $\Psi$ 
to describe the nodal excitation of quasiparticles in d-wave superconductors, 
and derived a low energy effective theory. 
They considered the coupling between the quasiparticles and fluctuating vortices of the system
by a gauge interaction. 
Then the low energy effective theory becomes 
the three- ( one for time and two for space ) dimensional two-flavor massless quantum electrodymanics 
( ${\rm QED}_{3}$ ).
The massless ${\rm QED}_{3}$ has a chiral symmetry; 
The Lagrangian is invariant under $\Psi\to e^{i\gamma_{5}\theta}\Psi$.  
It is a famous fact that the four-component ${\rm QED}_{3}$ dymanically 
generates a parity-conserving Dirac mass~[3]. 
Consulting on the field-theoretical result of ${\rm QED}_{3}$, 
they discussed the chiral symmetry breaking in their low-energy effective theory, 
and they observed that the chiral condensate 
$m_{dyn}\langle\bar{\Psi}\Psi\rangle$ ( $m_{dyn}$; the dynamical mass )
is an alternating spin density wave ( SDW ). 
Based on the result, they argued that the system ( d-wave superconductor with fluctuating vortices ) 
has an antiferromagnetic instability as the dynamical origin of ${\rm QED}_{3}$ model, 
and explained the reason of the existence of the antiferromagnetic order 
in the phase diagram of copper oxide superconductors.

The essential part of their discussions and conclusions, 
especially about the phenomenon of ${\rm D}\chi{\rm SB}$ 
can also be obtained by the following Lagrangian:
\begin{eqnarray}
\sum^{2}_{n=1}\bigl(\bar{\Psi}_{n}i\gamma^{\mu}\partial_{\mu}\Psi_{n}+G^{(3)}_{0}[(\bar{\Psi}_{n}\Psi_{n})^{2}+(\bar{\Psi}_{n}i\gamma_{5}\Psi_{n})^{2}]\bigr).
\end{eqnarray}
Here, we take the same definition of $\Psi_{n}$ as that of the ${\rm QED}_{3}$ model given by 
Franz-Tesanovic and Herbut. 
This model is simple, and at least for studying the ${\rm D}\chi{\rm SB}$,
the calculation is easier ( though we have to introduce a cutoff ) 
than the gauge theory, ${\rm QED}_{3}$.
It is clear from their logic, the ${\rm QED}_{3}$ model
can be applied to {\it all} d-wave superconductors
( not only to copper oxide but also to d-wave organic superconductors ).
Our four-fermi model (1) can also be applied to {\it all} d-wave superconductors. 
By introducing the local one-particle density matrix 
$Q(x)=-\langle\Psi(x)\bar{\Psi}(x)\rangle$, we proceed to perform
the group-theoretical classification for the order parameter developed from our theory~[4$\sim$7].
$Q(x)$ is a 4$\times$4 matrix, then we can expand it by 16-dimensional complete set of gamma matrices:
\begin{eqnarray}
Q &=& Q^{s}\hat{1}+Q^{V}_{\mu}\gamma^{\mu}+Q^{T}_{\mu\nu}\sigma^{\mu\nu}+Q^{A}_{\mu}\gamma_{5}\gamma^{\mu}+Q^{P}i\gamma_{5}.
\end{eqnarray} 
Here $S$, $V$, $T$, $A$ and $P$ denote the scalar, vector, tensor, axial vector and pseudoscalar, respectively. 
In fact, the dynamical mass discussed by Franz-Tesanovic and Herbut corresponds to the scalar density $Q^{S}$.
If we examine each component of the matrix $Q$ more in detail, 
we can discuss the possibility of the appearances of other types of order. 
Now we study the problem, and intend to publish our results elsewhere.

On the other hand, a low-energy effective theory for point-like-node p-wave superconductors 
( similar to the case of the ABM ( Anderson-Brinkman-Morel ) state )
becomes a two-dimensional ( one for time and one for space ) massless Dirac fermion model: 
The system has two Fermi points in a specific direction in momentum space, 
and quasiparticles are easily excited near the Fermi points. 
If we describe the low-energy long-wavelength excitation 
by $\psi_{R\sigma}(z)e^{ik_{F}z}+\psi_{L\sigma}(z)e^{-ik_{F}z}$, 
( here, $R$ denotes a right mover, $L$ denotes a left mover and $\sigma$ denotes a spin quantum number )
we will obtain a two-flavor massless Dirac fermion model:
\begin{eqnarray}
\sum_{\sigma}\bar{\psi}_{\sigma}i\gamma^{\mu}\partial_{\mu}\psi_{\sigma}.
\end{eqnarray}
Here, we take the definition of the two-component Dirac field 
as $\psi_{\sigma}=(\psi_{R\sigma}(z),\psi_{L\sigma}(z))$.
The gamma matrices are given 
by $\gamma_{0}=\sigma_{1}$, $\gamma_{1}=-i\sigma_{2}$, $\gamma_{5}=\gamma_{0}\gamma_{1}=\sigma_{3}$.
The effective Lagrangian (3) also has the chiral symmtery. 
If we consider a chiral invariant four-body contact interaction, 
its mathematical form is severely restricted. 
Add a chiral invariant interaction to the Dirac kinetic term (3), we get
\begin{eqnarray}
\sum^{2}_{n=1}\bigl(\bar{\psi}_{n}i\gamma^{\mu}\partial_{\mu}\psi_{n}+G^{(2)}_{0}[(\bar{\psi}_{n}\psi_{n})^{2}+(\bar{\psi}_{n}i\gamma_{5}\psi_{n})^{2}]\bigr).
\end{eqnarray}
Any continuous symmetry in one-dimension cannot be spontaneously broken~[8].
In the one-dimensional case, the non-Abelian bosonization procedure~[9$\sim$14] should be employed.
After incorporate the band multiplicity in our model,
our Hamiltonian will be decoupled to three sectors: 
$U(1)$ ( charge ), $SU(2)$ ( spin ) and $SU(N)$ ( orbital or band multiplicity ). 
Then we can write down the bosonized Hamiltonian in the Sugawara form~[15]:
\begin{eqnarray}
H &=& H_{U(1)}+H_{SU(2)}+H_{SU(N)}, \\
H_{U(1)} &=& 2\pi v_{charge}\int dx(:J_{R}J_{R}:+:J_{L}J_{L}:+G:J_{R}J_{L}:), \\
H_{SU(2)} &=& \frac{2\pi}{2+N}v_{spin}\sum^{3}_{a=1}\int dx(:J^{a}_{R}J^{a}_{R}:+:J^{a}_{L}J^{a}_{L}:-G:J^{a}_{R}J^{a}_{L}:), \\
H_{SU(N)} &=& \frac{2\pi}{2+N}v_{orb}\sum^{N^{2}-1}_{A=1}\int dx(:J^{A}_{R}J^{A}_{R}:+:J^{A}_{L}J^{A}_{L}:), 
\end{eqnarray}
In the expression given above, the spin-charge-orbital separation was occured.
By using the conformal field theoretical techniques with renormalization group approach~[12,13], 
we can predict that the excitation in each sector becomes gapless ( massless ) or gapful ( massive )~[13].  
Then we determine what kind of order ( CDW, SDW and "orbital wave" ) will emerge. 
For example, when the spectrum of the charge sector is massless, CDW will arise, 
while it is massive, CDW will not appear. 
It is clear from our discussion, this model can be applied to {\it all} systems 
which have point-like p-wave nodes. 
To examine the physics of CDW, Sakita et al. used the same Lagrangian with (4). 
They discussed the chiral symmetry of the Lagrangian to study the CDW~[16]. 
The most important point in our discussion is in the following logic: 
The excitation of p-wave superconductors will be described by the chiral invariant model, 
and when a kind of perturbation ( interaction between particles ) is applied, 
CDW or SDW may appear/disappear.
Because of the dimensionality of the nodal excitation in p-wave systems,
CDW and/or SDW can appear.   
Our context in this paper is different from that of Su and Sakita.

It should be emphasized that our theory, combined (1) with (4), can explain SDW, CDW and other possible
phases, while a phenomenological Landau-Ginzburg-type $SO(5)$ model introduced by Zhang~[17] can only explain
antiferromagnetic phase, superconducting phase and coexistense of them.

Let us consider various superconducting systems of real substances. 
Recently, some experiments found the existence and/or coexistence of CDW, SDW and other ordered state 
in some superconductors. 
For example, the coexistences of CDW and SDW 
in $({\rm TMTTF})_{2}{\rm Br}$, $({\rm TMTSF})_{2}{\rm PF}_{6}$ ( p- or f-wave superconductor ) 
and $\alpha -({\rm BEDT-TTF})_{2}{\rm MHg}({\rm SCN})_{4}$ ( non-pure s-wave ) were observed. 
The phese diagram of $({\rm BEDT-TTF})_{3}{\rm Cl}_{2}({\rm H_{2}O})_{2}$ has a CDW phase neighbor 
a superconducting phase.  
The importance of charge fluctuation with ferromagnetic spin fluctuation 
in ${\rm Sr_{2}RuO_{4}}$ ( p- or f-wave superconductor ) was pointed out by Takimoto~[18]. 
Kuroki et al. performed a theoretical investigation about the effect of the coexistence of CDW and SDW
in $({\rm TMTSF})_{2}{\rm PF}_{6}$~[19].  
Neighbor the superconducting phase of ${\rm UGe_{2}}$ ( p- or f-wave superconductor ),
there is a CDW/SDW coexistent phase. 
We recognize almost all of these substances are p- or f-wave superconductors. 
We speculate that the CDW phase or CDW/SDW coexistent phase may emerge 
by the mechanism of the generation of chiral mass in two-dimensional system, 
or by collective excitations of charge and spin in one-dimensinal system. 
We suppose the pairing symmetry of the superconducting phase 
in $({\rm BEDT-TTF})_{3}{\rm Cl}_{2}({\rm H_{2}O})_{2}$ is a p-wave type
( though there is no experimental report about it ).
We would like to make an argument that 
p-wave, d-wave and f-wave superconductors generally have the SDW/CDW instability.
To the contrary, s-wave superconductors do not have such kind of instability. 
Usually, the phase diagrams of p-wave, d-wave and f-wave superconductors have several ordered phases,
while the phase diagram of s-wave should become a simple one.
The chiral symmetry arises from the nodal structure of superconducting gap, 
and play the key-role in the coexistence/competition of various phases in phase diagrams of superconductors.

Finally, we wish to make a comment on the confinememt-deconfinement transition ( CDT ) in superconductors.
In quantum chromodynamics ( QCD ), quark confinement occurs at low-energy low-density state,
and the ${\rm D}\chi{\rm SB}$ is realized, dynamical mass is generated. 
At high-density state, the quark-deconfinement occurs and the color-superconductivity will be realized~[20,21]. 
Similar to this case, the confinement wil be realized at low-energy in ${\rm QED}_{3}$,
while deconfinement will occur at high-density state~[22].
There is a similarity between the phase diagram of copper oxide and that of QCD: 
We speculate SDW corresponds to ${\rm D}\chi{\rm SB}$ phase, 
while superconductivity corresponds to color-superconductivity. 
It is case that there are several similarities between ${\rm QED}_{3}$ and ${\rm QCD}_{4}$. 
Therefore, there is a possibility to understand the phase diagram of copper oxide by the concept of CDT. 
We suppose both the ${\rm D}\chi{\rm SB}$ and CDT are universal phenomena 
in various condensed matter.

\end{document}